\documentclass{aastex631}
\usepackage{makecell}
\usepackage{amsmath}
\usepackage{cases}

\begin{document}{\nolinenumbers}

\title{Detection of the extended $\gamma$-ray emission around supernova remnant DA 530 with {\em Fermi}-LAT}

\author[0000-0001-5135-5942]{Yuliang Xin}
\author[0009-0004-5450-8237]{Xiaolei Guo}
\affiliation{School of Physical Science and Technology, Southwest Jiaotong University, Chengdu 610031, China\\
\href{mailto:ylxin@swjtu.edu.cn}{ylxin@swjtu.edu.cn}}



\begin{abstract}
We report the extended GeV $\gamma$-ray emission around the high Galactic latitude supernova remnant (SNR) DA 530 with the PASS 8 data recorded by the Fermi Large Area Telescope ({\em Fermi}-LAT).
The $\gamma$-ray spectrum in the energy range of 100 MeV - 1 TeV follows a power law model with an index of 2.23. 
The much more extended $\gamma$-ray emission than the radio shell of DA 530 and the spatial coincidence with the molecular cloud suggest that the $\gamma$-ray emission could be originated from the hadronic process, where the high energy protons are accelerated in and escaped from the shock of DA 530.
With a steady-state injection model of protons, the $\gamma$-ray spectrum can be well fitted with the typical Galactic value for diffusion coefficient and the low energy content of the total escaped protons.
\end{abstract}



\section{Introduction} 
\label{sec:intro}
Supernova remnants (SNRs) are widely believed to be the dominant accelerators of Galactic cosmic rays (CRs). And CRs can be accelerated by the high speed shock of SNR with the mechanism of diffusive shock acceleration \citep[DSA;][]{1987PhR...154....1B}. 
The highest-energy CRs are expected to escape from the shock due to the absence of the self-generated magnetic turbulence \citep{2005A&A...429..755P,2011MNRAS.415.3434F}.
If there are some molecular clouds (MCs) in the vicinity of SNRs, these MCs could be illuminated by the escaped protons and produce the intense $\gamma$-ray emission with the hadronic process, i.e., the $\gamma$-ray emission are believed to be from the decay of neutral pions produced in inelastic collisions between escaped protons and the dense gas in MCs.
And several sources have been detected in this perspective, e.g. W28 \citep{2008A&A...481..401A,2018ApJ...860...69C}, W44 \citep{2012ApJ...749L..35U}, SNR G15.4+0.1 \citep{2023ApJ...945...21L}, SNR G45.7-0.4 \citep{2021ApJ...923..106Z}, etc.
Especially in the case of W28, the GeV and TeV $\gamma$-ray emission are detected in three distinct regions, which are spatially coinciding with MCs offset from the shock of SNR \citep{2008A&A...481..401A,2018ApJ...860...69C}.
Searching for the $\gamma$-ray emission from such associations could be helpful to explore the propagation properties of high energy particles escaped from SNRs and afterwards travelled through the interstellar medium (ISM), which further provides a key ingredient to establishing the connection between SNRs and the origin of Galactic CRs.

DA 530, also known as G93.3+6.9, is a high Galactic latitude SNR \citep{1976A&A....51..151R}.
And based on the radio observations, DA 530 is classified to be a shell-type SNR with a bilateral morphology \citep{1999ApJ...527..866L}.
The radio radiation of DA 530 has extremely high polarization percentage, reaching more than 50\%, which could be interpreted by the well-ordered magnetic field across the whole remnant \citep{1980A&A....92...57H,1984A&A...131..196L}.
The X-ray emission of DA 530 first detected by {\em ROSAT} is extremely faint with a centrally brightened morphology \citep{1999ApJ...527..866L}.
By re-analysing the {\em Chandra} data, \citet{2007ApJ...670.1142J} found a small-scale hard X-ray feature near the centre of the remnant, which is argued to be a pulsar wind nebula (PWN) associated with SNR.
And the age of DA 530 was suggested to be $\sim$ 5000 yrs based on the canonical blast wave model from \citet{1959sdmm.book.....S}.
However, the subsequent {\em XMM}-Newton observations detected a large extended source (XMM J205314.4+551528) in the radio bright southeast (SE) rim of DA 530 \citep{2008AdSpR..41..407B},
and the authors explained it to be the PWN associated with DA 530.
The recent {\em Suzaku} data analysis of the SE rim of DA 530 confirmed the results of {\em XMM}-Newton, which suggested that the PWN scenario can not be ruled out \citep{2022AdSpR..69.2342D}. 
Nonetheless, the searching for the pulsars or compact central sources associated with DA 530 in the radio and X-ray bands show the null results \citep{1998A&A...331.1002L,1999ApJ...527..866L,2004ApJS..153..269K, 2019A&A...623A..90S}.
The distance of DA 530 is not very clear so far, which was first derived to be 6.9 $\pm$ 2.2 kpc using the empirical surface brightness-diameter ($\Sigma$ - {\em D}) relation of SNRs \citep{1976A&A....51..151R}. 
Then an updated distance of 2 - 5 kpc was given by \citet{1980A&A....92...57H}.
Based on the neutral hydrogen (HI) observations with Dominion Radio Astrophysical Observatory Synthesis Telescope (DRAO-ST), 
\citet{1999ApJ...527..866L} reported that DA 530 lies within a shell of HI, possibly created by an earlier stellar wind of the progenitor. 
And the distance of it is estimated to be 1.0 - 3.5 kpc.
Subsequently, \citet{2003ApJ...598.1005F} derived a distance of 2.2 $\pm$ 0.5 kpc using the updated method for the absorption column density.
Using the data from DRAO-ST and National Radio Astronomy Observatory Very Large Array (NRAO-VLA), \citet{2022ApJ...941...17B} observed the absorption by intervening HI of the polarized emission from DA 530, 
and concluded that the distance of DA 530 is 4.4$^{\rm +0.4}_{\rm -0.2}$ kpc.
Using the Seoul Radio Astronomy Observatory (SRAO) 6-m telescope CO observations, \citet{2012Ap&SS.342..389J} detected CO emission at -6 to +5 km s$^{\rm -1}$ at the northeast boundary of DA 530, which shows that there is a large diffuse molecular cloud.

In this work, we report the detection of the extended $\gamma$-ray emission around SNR DA 530, with the PASS 8 data recorded by {\em Fermi}-LAT. 
The data analysis method and results are shown in Section 2, including the spatial and spectral analyses.
The observations of molecular cloud around DA 530 is presented in Section 3.
And the discussion of the potential origin of the $\gamma$-ray emission is shown in Section 4, followed by the summary in Section 5.

\section{{\em Fermi}-LAT Data Analysis}
\label{fermi}

\subsection{Data Reduction}
\label{data reduction}
{\em Fermi}-LAT is a pair-conversion $\gamma$-ray telescope that is sensitive to photon energies greater than 20 MeV. The LAT has continuously monitored the sky since 2008 and scans the entire sky every 3 hr \citep{2009ApJ...697.1071A}. Note that the latest released Pass 8 data set has significant improvements in comparison with the former ones, including an enhanced effective area, especially in the low energy range and better point-spread function (PSF)\footnote{https://www.slac.stanford.edu/exp/glast/groups/canda/lat$\_$Performance.htm}.
And in the following analysis, we select the latest Pass 8 version of {\em Fermi}-LAT data recorded from August 4, 2008 (Mission Elapsed Time 239557418) to August 4, 2022 (Mission Elapsed Time 681264005) with ``Source'' event class (evclass = 128 \& evtype = 3) to analyse the $\gamma$-ray emission around DA 530.
The region of interest (ROI) is a $20^\circ \times 20^\circ$ square region centered at the position of DA 530 \citep[R.A. = $313^{\circ}\!.14$, decl. = $55^{\circ}\!.36$;][]{1976A&A....51..151R}.
And in order to reduce the contamination from Earth Limb, the events with zenith angle larger than $90^\circ$ are excluded.
We adopt the events with energy range of 100 MeV - 1 TeV for the spectral analysis. While for the spatial analysis, the events in the energy range of 1 GeV - 1 TeV is selected considering the impact of PSF of {\em Fermi}-LAT.
The data are analyzed using the standard {\it Fermi ScienceTools} \footnote {http://fermi.gsfc.nasa.gov/ssc/data/analysis/software/} 
with the instrumental response function (IRF) of ``P8R3{\_}SOURCE{\_}V3''.
The binned likelihood analysis method with {\em gtlike} is used to fit the data.
To model the Galactic and isotropic diffuse background emissions,
{\tt gll\_iem\_v07.fits} and {\tt iso\_P8R3\_SOURCE\_V3\_v1.txt}
\footnote {http://fermi.gsfc.nasa.gov/ssc/data/access/lat/BackgroundModels.html} are adopted.
All sources in the incremental version of the fourth {\em Fermi}-LAT source catalog \citep[4FGL-DR3;][]{2020ApJS..247...33A, 2022ApJS..260...53A} within a radius of $20^\circ$ from the ROI center
and the two components of the diffuse background, are included in the source model, which is generated by the user-contributed software 
{\tt make4FGLxml.py}\footnote{http://fermi.gsfc.nasa.gov/ssc/data/analysis/user/}.
During the likelihood analysis, the normalizations and the spectral parameters of all sources within $7^{\circ}$ to the center of ROI, together with the normalizations of the two components of the diffuse background, are set to be free.

\subsection{Spatial Analysis}
\label{Spatial}

\begin{figure*}[!htb]
	\centering
    \includegraphics[width=0.48\textwidth]{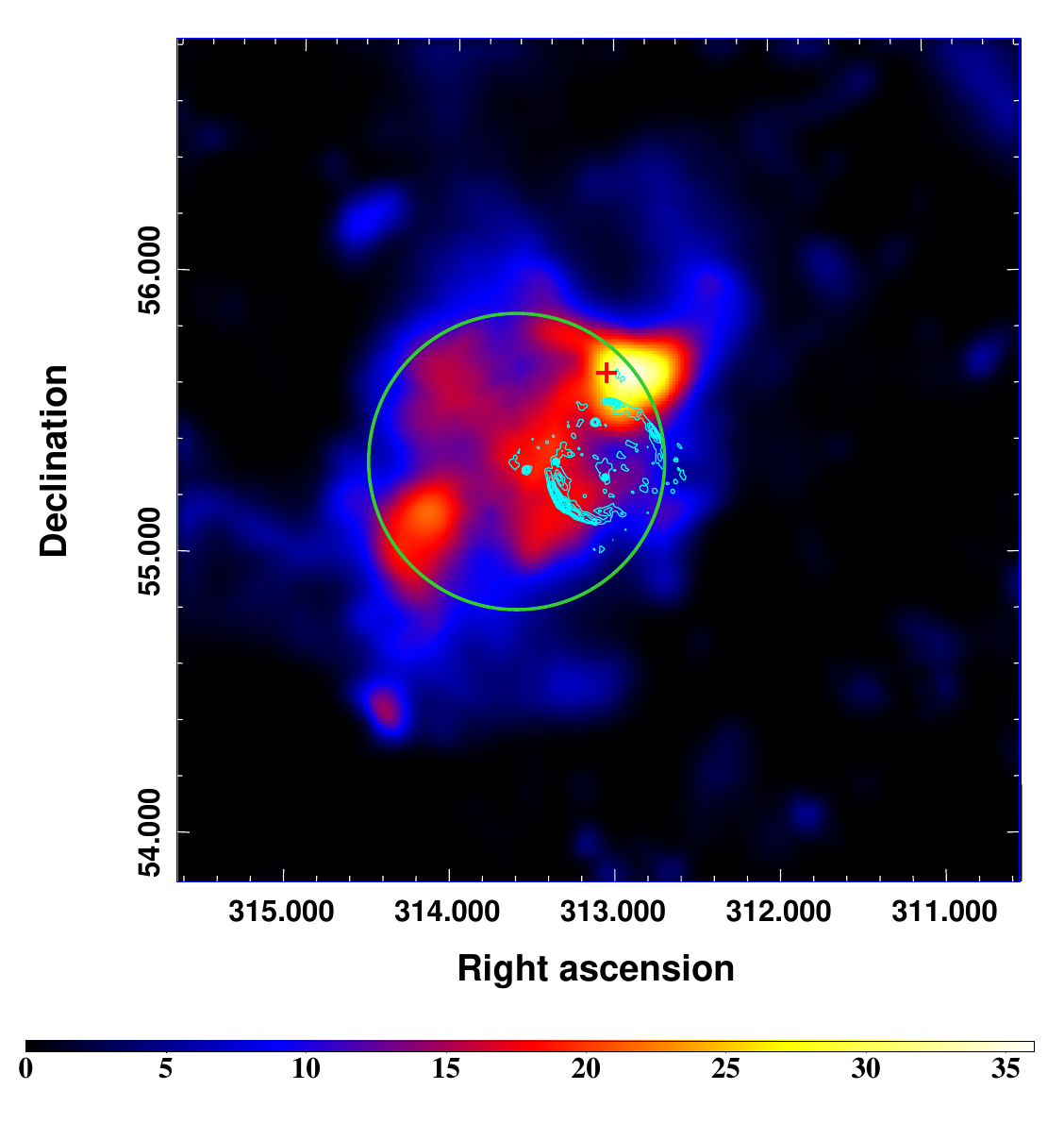}
    \includegraphics[width=0.48\textwidth]{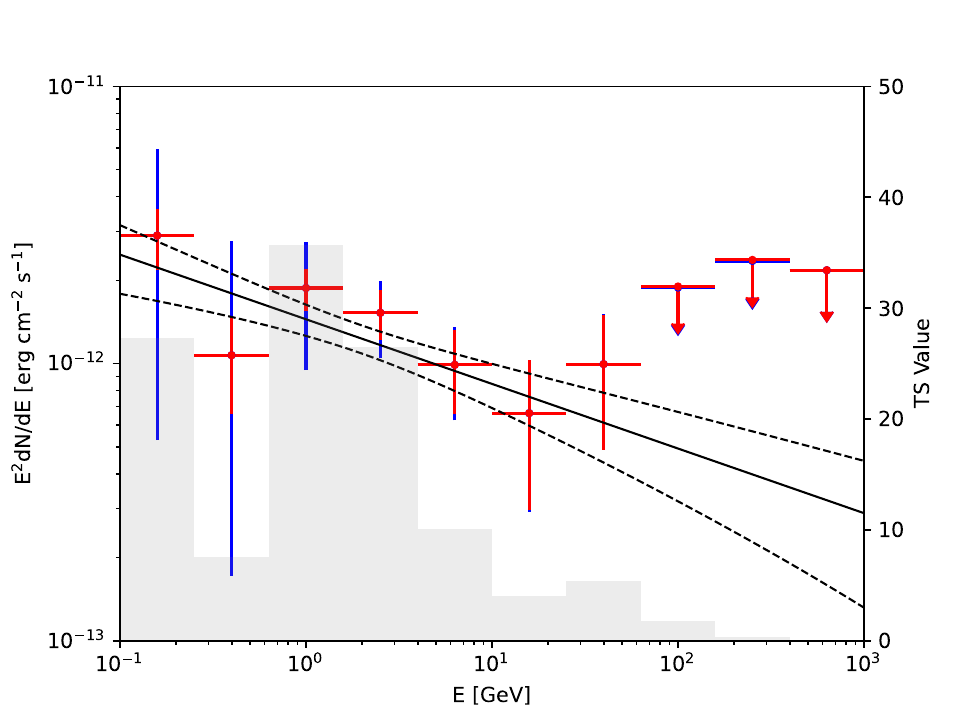}
	\caption{Left: $3^{\circ}\!.0$ $\times$ $3^{\circ}\!.0$ TSmap in the energy range of 1 GeV - 1 TeV. The red cross shows the position of 4FGL J2051.1+5539 in 4FGL-DR3 catalog. And the green solid circle marks the best-fit 68\% containment radius of the uniform disk for the spatial template of SrcT. The radio image of SNR DA 530 at 1.4 GHz is shown as the cyan contours. Right: SED of SrcT in the energy range of 100 MeV - 1 TeV with the gray histogram shown as the TS value for each energy bin. The red error bars show the statistical errors and the sums of the statistical and systematic errors calculated by $\sigma$ = $\sqrt{\sigma_{\rm stat}^2 + \sigma_{\rm syst}^2}$ are marked by the blue error bars. The arrows indicate the 95\% upper limits for the energy bin with TS value of SrcT smaller than 5.0. The black solid and dashed lines show the global best-fit power law spectrum and its 1$\sigma$ statistic error in the energy range of 100 MeV - 1 TeV.}
	\label{fig1:tsmap}
\end{figure*}

In the region of DA 530, a $\gamma$-ray point source (4FGL J2051.1+5539) is listed in the 4FGL-DR3 catalog, which has no identified counterpart \citep{2022ApJS..260...53A}.
First, we create a $3^{\circ}\!.0$ $\times$ $3^{\circ}\!.0$ Test Statistic (TS) map by subtracting the emission from the sources and backgrounds (except 4FGL J2051.1+5539) in the best-fit model with {\em gttsmap}, which is shown in the left panel of Figure \ref{fig1:tsmap}.
At the center of ROI, a significant $\gamma$-ray excess (labelled as SrcT) is found around SNR DA 530.
For the spatial template of SrcT, we first treat it to be a point source, and the best-fit position is given to be R.A. = $312^{\circ}\!.983 \pm 0^{\circ}\!.029$, decl. = $55^{\circ}\!.666 \pm 0^{\circ}\!.029$ by adopting  {\tt Fermipy}, a {\tt PYTHON} package that automates analyses with {\it Fermi ScienceTools} \citep{2017ICRC...35..824W}.
The TS value of SrcT as a point source is 33.46 with the new coordinate.
Then we carried out a spatial extension test for the $\gamma$-ray emission of SrcT using an uniform disk and a two-dimensional (2D) Gaussian template with {\tt Fermipy}.
And the best-fit central positions and extensions of the spatial templates are listed in Table \ref{table:spatial}, together with the corresponding TS value of SrcT and the maximum likelihood values.
We compared the overall maximum likelihood of the extended template ($\rm \mathcal{L}_{\rm ext}$) with that of the point source model ($\rm \mathcal{L}_{\rm pt}$), and the significance of the extended model is defined to be $\rm {TS_{\rm ext}}=2(\ln\mathcal{L}_{\rm ext}-\ln\mathcal{L}_{\rm pt})$.
\citet{2012ApJ...756....5L} suggests that a source can be assessed to be significantly extended if $\rm {TS_{\rm ext}} > 16$.
The different maximum likelihood values of the different templates show that an uniform disk can best fit the $\gamma$-ray emission from SrcT.
And the central position and the 68\% containment radius of the uniform disk are fitted to be R.A. = $313^{\circ}\!.611$, decl. = $55^{\circ}\!.344$ and R$_{\rm 68}$ = $0^{\circ}\!.527$, respectively.
The value of $\rm {TS_{\rm ext}}$ between the uniform disk model and point source model is calculated to be 26.2, corresponding to $\sim$5.1$\sigma$ extension with one additional degree of freedom (dof).
With the uniform disk template, the TS value of SrcT is fitted to be 58.75 in the energy range of 1 GeV - 1 TeV, corresponding to a significance level of $\sim$6.7 $\sigma$ with five degrees of freedom.

\begin{table}[!htb]
\centering
\caption {Spatial Analysis for SrcT in the energy range of 1 GeV - 1 TeV}
\begin{tabular}{ccccccc}
\hline \hline
Spatial Model    &  R.A., decl.    & TS Value    & Degrees of Freedom   & -log(Likelihood)\\
\hline
Point Source      & $312^{\circ}\!.983 \pm 0^{\circ}\!.029$, $55^{\circ}\!.666 \pm 0^{\circ}\!.029$  &  33.46  & 4   & -182258.99 \\ 
\hline
Uniform disk  & \makecell[c]{$313^{\circ}\!.611 \pm 0^{\circ}\!.043$, $55^{\circ}\!.344 \pm 0^{\circ}\!.049$, \\ R$_{\rm 68}$ = $0^{\circ}\!.527 ^{+0^{\circ}\!.028}_{-0^{\circ}\!.029}$}  & 58.75  & 5  & -182272.08 \\
\hline
2D-Gaussian   & \makecell[c]{$313^{\circ}\!.499 \pm 0^{\circ}\!.100$, $55^{\circ}\!.415 \pm 0^{\circ}\!.102$, \\ R$_{\rm 68}$ = $0^{\circ}\!.567 ^{+0^{\circ}\!.101}_{-0^{\circ}\!.080}$}  & 53.93  & 5  & -182269.65\\
\hline
\end{tabular}
\label{table:spatial}
\end{table}   

\subsection{Spectral Analysis}
\label{Spectral}

To investigate the $\gamma$-ray spectrum of SrcT, the global likelihood analysis is performed in the energy range from 100 MeV to 1 TeV with the spatial template of an uniform disk. 
The spectrum of SrcT can be well described by a power law (PL) model. 
The spectral index and the integral photon flux in the energy range of 100 MeV - 1 TeV are fitted to be 2.23 $\pm$ 0.09 and $(1.25 \pm 0.23)\times10^{-8}$ photon cm$^{-2}$ s$^{-1}$, respectively.
And we also adopted a log-parabola model (LPb) to test the spectral curvature of SrcT. The variation of the maximum likelihood values between PL and LPb models is only $\rm {TS_{\rm curve}}=2(\ln\mathcal{L}_{\rm LPb}-\ln\mathcal{L}_{\rm PL})=2.1$, which suggests no significant curvature for the $\gamma$-ray spectrum of SrcT.
With the PL model for SrcT, we divide the data into ten logarithmically equal energy bins from 100 MeV to 1 TeV and repeat the likelihood fitting to give the spectral energy distribution (SED) of SrcT.
For the likelihood analysis, only the spectral normalizations of sources within $7^\circ$ from SrcT are left free, 
together with the normalizations of the two components of the diffuse background.
And the spectral indices of these sources are fixed to be the best-fit values in the global likelihood analysis.
For the energy bin with a TS value of SrcT smaller than 5.0, an upper limit with a 95\% confidence level is calculated. 
For the SED, we also estimated the systematic errors due to the Galactic diffuse emission by changing the best-fit normalization of the Galactic diffuse model artificially by $\pm$6\% \citep{2010ApJ...714..927A}.
And the sums of the statistical and systematic errors are calculated by $\sigma$ = $\sqrt{\sigma_{\rm stat}^2 + \sigma_{\rm syst}^2}$ for each energy bin.
The SED of SrcT is shown in the right panel of Figure \ref{fig1:tsmap}, which is also consistent with the global fitting of the power law model.

\section{CO Observation}
\label{CO}

\begin{figure*}[!htb]
    \centering
    \includegraphics[width=0.5\textwidth]{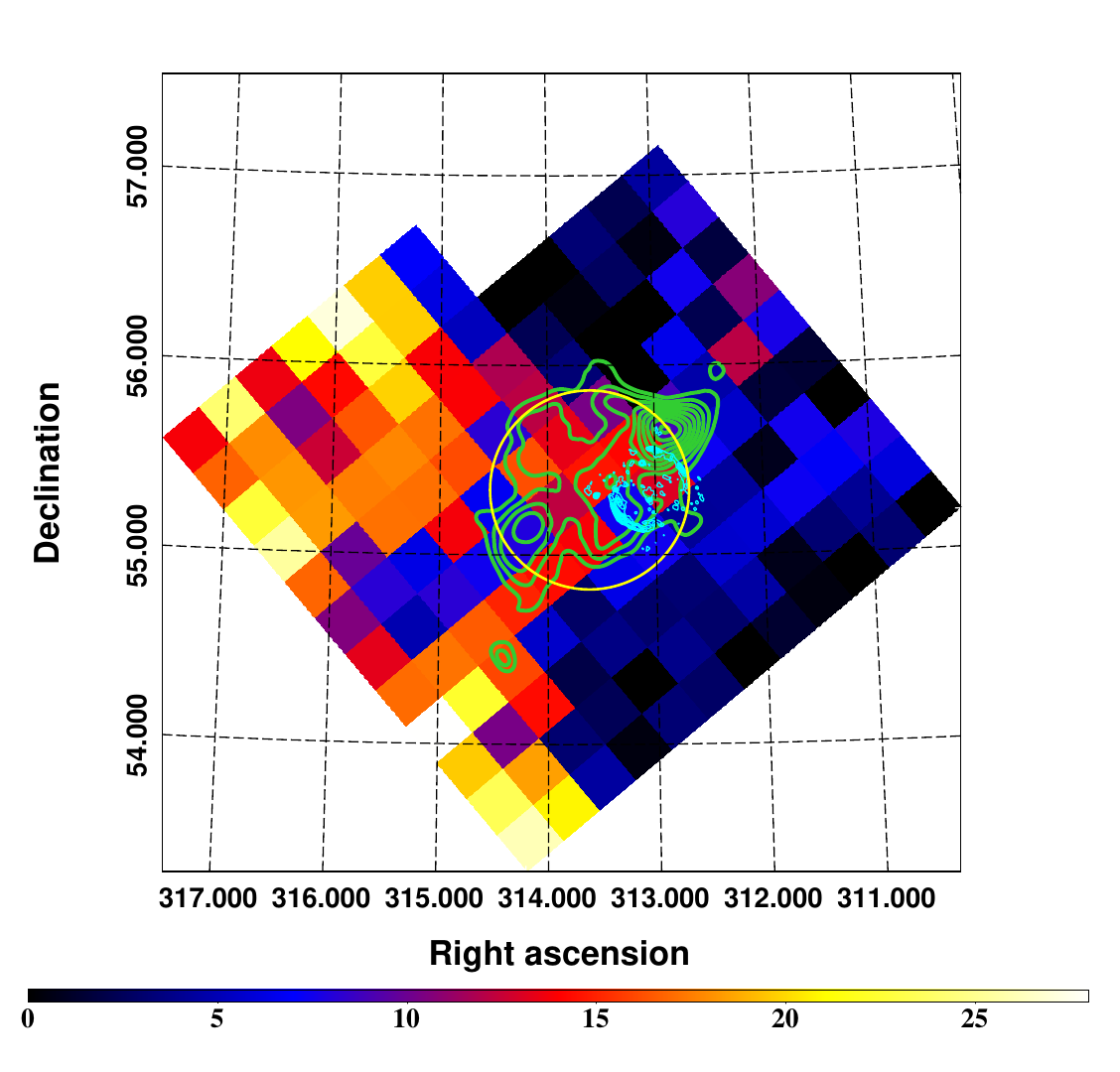}
    \caption{Integrated CO emission intensity (K km s$^{-1}$) around DA 530 in the velocity range of -6 $\sim$ +5 km s$^{-1}$. 
The 68\% containment radius of the extended $\gamma$-ray emission of SrcT is marked by the yellow solid circle. The green and cyan contours show the $\gamma$-ray morphology of SrcT and the radio image of SNR DA 530 at 1.4 GHz, as shown in the left panel of Figure \ref{fig1:tsmap}.}
    \label{fig2-COmap}
\end{figure*}

With the CO observations from SRAO, \citet{2012Ap&SS.342..389J} claimed a large diffuse molecular cloud surrounding the northeast boundary of DA 530.
However, the size of the $\gamma$-ray emission detected here is much larger than that of their CO observation. Therefore, we try to search for the components of molecular cloud in a large extend using the data from the CfA 1.2m millimeter-wave telescope to understand the origin of the $\gamma$-ray excess \citep{2001ApJ...547..792D}.
And we found that in the $\gamma$-ray emission region, the velocity distribution of the CO content shows a clear excess in the velocity range of -6 $\sim$ +5 km s$^{\rm -1}$ as shown in Figure \ref{fig2-COmap}, which is consistent with the range in \citet{2012Ap&SS.342..389J}.
Adopting the standard Galactic rotation model \citep{2016ApJ...823...77R,2019ApJ...885..131R}, the velocity interval corresponds to a kinetic distance of $\sim$1.6 kpc.
And considering the systematic uncertainties due to the rotation curve, the derived distance is close to the value of 2.2 $\pm$ 0.5 kpc derived by \citet{2003ApJ...598.1005F} using the updated method for the absorption column density, which is also much lower that the result of 4.4 kpc in \citet{2022ApJ...941...17B}.

We have then estimated the mass content of the molecular material in the $\gamma$-ray emission region with 
\begin{equation}
 M= \mu m_\mathrm{H} d^2 \Omega_\mathrm{px} X_\mathrm{CO} {\sum_\mathrm{px}} W_\mathrm{CO}
\end{equation}
\\
Here the mean molecular weight $\mu$ is adopted to be 2.8 assuming a relative helium abundance of 25\%.
$m_\mathrm{H}$ is the mass of the Hydrogen nucleon, and the distance is adopted to be d = 2.2 kpc \citep{2003ApJ...598.1005F}. 
$\Omega_\mathrm{px}$ is the solid angle subtended for each pixel in the map shown as in Figure \ref{fig2-COmap}.
And the value of the conversion factor of $X_\mathrm{CO} = 2\times10^{20}$ cm$^{-2}$ (K km s$^{-1})^{-1}$ is adopted here \citep{2013ARA&A..51..207B}.
$\sum_\mathrm{px}$W$_\mathrm{CO}$ is calculated by summing the map content of each pixel in the desired sky region and velocity range.
For the region of $0.527\degr$ sky integration radius, the total mass of molecular clouds in the region within the 68\% containment radius of the extended $\gamma$-ray source is estimated to be $\sim$ $2.6\times 10^{5}d_{2.2}^{2} M_{\odot}$.
And the corresponding average gas number density is about $\rm{n_{\rm gas}}$ = 300 cm$^{-3}$ by assuming a spherical geometry of the gas distribution.

\section{Discussion}
\label{discussion}

The {\em Fermi}-LAT data analysis above shows an extended $\gamma$-ray source SrcT, which locates around the high Galactic latitude SNR DA 530.
And the GeV $\gamma$-ray spectrum of SrcT can be described by a power law model with an index of 2.23 $\pm$ 0.09.
To determine the origin of the $\gamma$-ray emission, we also searched for the CO observation, and SrcT is spatially consistent with the molecular clouds, which is located outside the radio shell of DA 530.
The spatial coincidence between the $\gamma$-ray emission and the molecular clouds suggests that the $\gamma$-ray emission from SrcT could be produced by the hadronic $\pi^0$ decay originating from the interaction between the molecular gas and high energy protons, which are accelerated in and escaped from the shock of DA 530.
Such scenario is similar to the origin of the $\gamma$-ray emission around SNR G15.4+1.0 \citep{2023ApJ...945...21L} and the SNR associated with PSR J0837-2454\citep{2023ApJ...951..142Z}.

To explain the $\gamma$-ray emission from SrcT, we assume the steady-state injection of protons into an uniform emission region.
And the injection time is adopted to be the age of DA 530 with T = 5000 yrs from \citet{2007ApJ...670.1142J}.
The spectrum of injected protons is adopted to be a power law with an exponential cutoff:
\begin{eqnarray}
Q(E) = {Q_0} E^{-\Gamma} \exp \left(- \frac{E}{E_{\rm p, cut}} \right).
\label{eq:p_spectra}
\end{eqnarray}
\\
Here the spectral index is suggested to be $\Gamma = 2.0$, which is consistent with the radio spectral index of DA 530 of $\alpha$ = 0.45 $\pm$ 0.04 by assuming the same spectral index for electrons and protons accelerated by the shock of SNR.
The cutoff energy of protons can not be well constrained and is first adopted to be the energy of the cosmic ray knee with $E_{\rm p, cut}$ = 3 PeV.
And the total energy of the injected protons is assumed to be $\rm W_{\rm p,inj}$ = $\rm \eta E_{\rm SN}$, where $\eta$ is the fraction of the kinetic energy of DA 530 converted into the escaped proton energy, and the the kinetic energy of DA 530, $\rm E_{\rm SN}$, is adopted to be a typical value of 10$^{\rm 51}$ erg.

By integrating the Eq. 16 in \citet{2012MNRAS.419..624T} on the variable radius over a sphere with radius R,
the escaped proton spectrum within the $\gamma$-ray emission region can be derived as \citep{1996A&A...309..917A,2012MNRAS.419..624T}:
\begin{equation}
    N_p(E,T)=\frac{Q(E)}{4 \pi D(E) T} \int_{0}^{R} 4\pi r dr \; {\rm erfc} \left[\frac{r}{\sqrt{4 D(E) T}}\right] \label{equation:3}
\end{equation}
\\
Here, the diffusion coefficient of protons is set to be spatially constant and energy-dependent with $D(E)=\chi D_0(E/E_0)^\delta$, 
where $D_0=3\times 10^{28}$ cm$^2$ s$^{-1}$ at $E_0=10$ GeV and $\chi$ = 1.0 corresponds to the typical value of Galactic diffusion coefficient \citep{2013A&ARv..21...70B}.
The value of $\delta$ is adopted to be 1/3 or 1/2, which corresponds to the Kolmogorov turbulence or Kraichnan turbulence for the diffusion coefficient \citep{2006ApJ...642..902P}.
And with the distance of 2.2 kpc and 68\% containment radius of $ 0^{\circ}\!.527$ for SrcT,
the physical radius of the $\gamma$-ray emission region is estimated to be R = 20.2 pc.
For an injected spectrum given by $Q(E) \propto E^{-\Gamma}$ and $D(E) \propto E^\delta$, the spectrum of escaped protons, $N_p(E)$, approximately follows $N_p(E) \propto E^{-(\Gamma+\delta)}$ in the high energy, where the diffusion radius of protons defined as $\sqrt{4D(E)T}$ is larger than the size of the emission region R.
The typical values of $\eta$, $\chi$ and $\delta$ are selected to calculate the different spectra of escaped protons in the $\gamma$-ray emission region, and the corresponding $\gamma$-ray fluxes are calculated using the $naima$ package by adopting the ambient gas density of $\rm{n_{\rm gas}}$ = 300 cm$^{-3}$ \citep{naima}.

\begin{figure*}[!htb]
	\centering
    \includegraphics[width=0.5\textwidth]{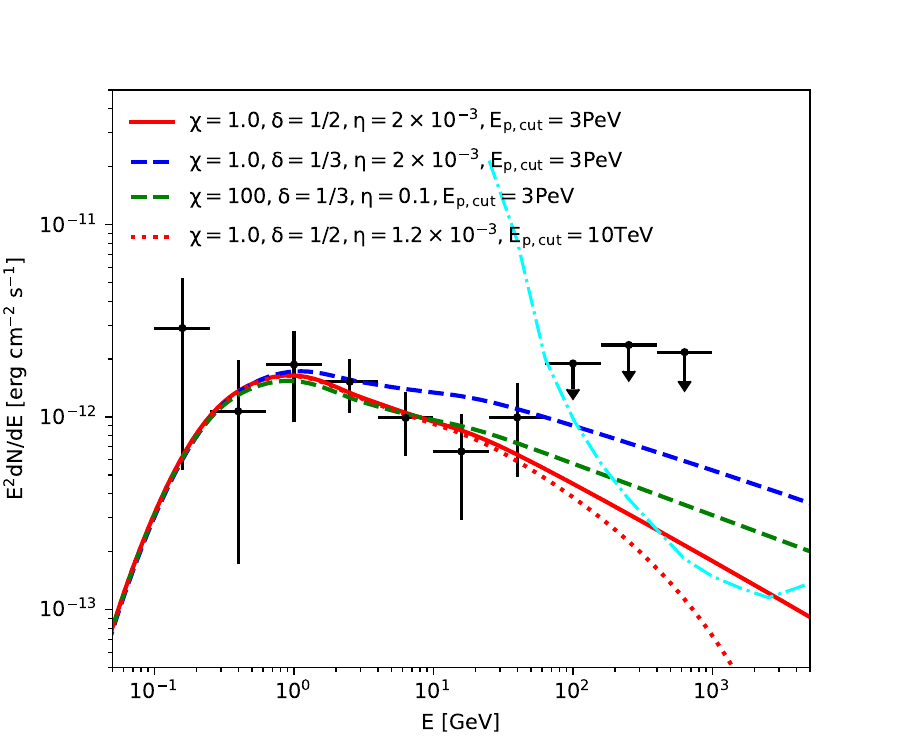}
	\caption{Modeling of the $\gamma$-ray spectra with the hadronic escaping models. The red solid, green and blue dashed lines indicate the scenarios with the different values for $\chi$, $\delta$ and $\eta$ and the cutoff energy of $E_{\rm p, cut}$ = 3 PeV. And the red dotted line shows the scenario with $E_{\rm p, cut}$ = 10 TeV, as shown in the legend. The cyan dot–dashed line shows the differential sensitivity of CTA-North \citep[50 hr;][]{2019scta.book.....C}.}
\label{Fig:Model}
\end{figure*}

Figure \ref{Fig:Model} shows the resulting hadronic $\gamma$-ray spectra with the different parameters.
The spectra with the typical value of Galactic diffusion coefficient, $\chi$ = 1.0, could explain the observational GeV data. 
And the total energy of injected protons above 1 GeV is fitted to be about 2.0 $\times$ 10$^{\rm 48}$ erg. 
which is lower than the estimated value by assuming the total acceleration efficiency of 5\% - 10\% and the typical kinetic energy of 10$^{\rm 51}$ erg for SNR \citep{2013A&ARv..21...70B}.
Such result could be attributed to the possible low kinetic energy of DA 530 with $\sim$10$^{\rm 49}$ erg \citep{2007ApJ...670.1142J}.
And it also could mean that the bulk of accelerated particles are still trapped inside the remnant, 
and the non-significant $\gamma$-ray emission within the shell of DA 530 shown in the left panel of Figure \ref{fig1:tsmap} could be explained by the low gas density inside SNR, which is also consistent with the evolution environment of a stellar wind bubble for DA 530 \citep{1999ApJ...527..866L}.
The total energies of escaped protons above 1~GeV in the $\gamma$-ray emission region are calculated to be 1.1 and 1.3 $\times$ 10$^{47}$ ($n_{\rm gas}$/300~${\rm cm^{-3}}$)$^{-1}$~erg for $\delta$ = 1/2 and $\delta$ = 1/3, respectively.
To increase the total energy of injected protons, one needs to increase the diffusion coefficient to make the $\gamma$-ray spectrum not be changed significantly.
By fixing the value of $\rm W_{\rm p,inj}$ = 10$^{\rm 50}$~erg, $\eta$ = 0.1, the diffusion coefficient needs to be about two orders of magnitude higher than the typical Galactic value. And the estimated total energy of escaped protons in the $\gamma$-ray emission region is about 1.2$\times$10$^{47}$ ($n_{\rm gas}$/300~${\rm cm^{-3}}$)$^{-1}$~erg.
In addition, we also decreased the cutoff energy of protons to estimate the different escaping models. And the allowed minimum value of cutoff energy is about 10 TeV, shown as the red dotted line in Figure \ref{Fig:Model} with the total energy of escaped protons in the $\gamma$-ray emission region of 1.1$\times$10$^{47}$ ($n_{\rm gas}$/300~${\rm cm^{-3}}$)$^{-1}$~erg, which expects a much lower flux in the TeV band.

Moreover, we also considered the the minimum energy of protons that can escape from the SNR with a simple but reasonable approach of $E_{\rm esc} = E_{\rm max}(t/t_{\rm sed})^{-\alpha}$ \citep{2009MNRAS.396.1629G,2011MNRAS.410.1577O,2012MNRAS.419..624T}, which assuming that the escape of the highest energy particles starts at the onset of the Sedov phase of SNR. Here the maximum energy of protons, $E_{\rm max}$, is assumed to be equal to $E_{\rm p, cut}$ with 3 PeV, and $t_{\rm sed}$ = 500 yrs \citep{2012MNRAS.419..624T}. SNR DA 530 with the age of 5000 yrs is suggested to be in the Sedov phase \citep{2007ApJ...670.1142J}, and the values of $E_{\rm esc}$ are calculated to be in the range of 30 TeV - 2 PeV for the different values of $\alpha$ from 0.2 to 2.0 \citep{2012MNRAS.419..624T}. And such energy range of escaping protons is in accord with the needed one that explaining the GeV $\gamma$-ray spectrum.

\section{Summary}
\label{summary}

In this work, we analyzed the GeV $\gamma$-ray emission in the field of SNR DA 530, using 14 years of {\em Fermi}-LAT data, and found an extended $\gamma$-ray source, SrcT, around DA 530, which can be described by an uniform disk template.
The GeV $\gamma$-ray spectrum of SrcT can be fitted by a power law model with an index of 2.23 $\pm$ 0.09.
The size of the $\gamma$-ray emission region is much larger than that of the radio shell of DA 530. 
Based on the CO observation from the CfA 1.2m millimeter-wave telescope, we found that the molecular cloud component is spatially consistent with the $\gamma$-ray emission region.
Considering the much more extended $\gamma$-ray emission and the spatial coincidence with the molecular cloud, the $\gamma$-ray emission of SrcT is suggested to be from the hadronic $\pi^0$ decay due to the inelastic collisions between the cloud and the high energy protons, which are accelerated in and escaped from the shock of DA 530.
With the assumption of steady-state injection of protons, the $\gamma$-ray spectrum can be well explained by the model with the typical Galactic value for the diffusion coefficient ($\chi = 1.0$). 
The total energy of the injected protons needs to be much lower, which can be explained by the low kinetic energy of DA 530 or the assumption that the bulk of accelerated particles are still trapped inside the remnant.
The hadronic $\gamma$-ray spectra with the different parameters expect the different fluxes in the TeV band. And the potential detection by the Cherenkov Telescope Array in the northern
hemisphere \citep[CTA-North;][]{2019scta.book.....C} in the future could help to test the different models and to constrain the escaping proton energy.
Moreover, the high-resolution observations for molecular cloud around DA 530 are necessary to clearly identify the origin of the extended $\gamma$-ray emission.

\hspace*{\fill} \\
We would like to thank the anonymous referee for very helpful comments, which help to improve the paper. 
This work is supported by the National Natural Science Foundation of China under the grants 12103040, 12147208 and U1931204, and the Natural Science Foundation for Young Scholars of Sichuan Province, China (No. 2022NSFSC1808).

\bibliography{sample631}{}
\bibliographystyle{aasjournal}



\end{document}